\documentclass[11pt]{article}
\usepackage{jheppub}
\usepackage{amsmath,amssymb,amsfonts,graphicx,slashed,amsthm,mathtools,upgreek, enumerate, tensor, subfig}
\usepackage[dvipsnames]{xcolor}
\usepackage{arydshln}

\usepackage{comment}
\usepackage{hyperref}
\usepackage[utf8]{inputenc}
\usepackage[titletoc]{appendix}

\numberwithin{equation}{section} 
\allowdisplaybreaks

\allowdisplaybreaks

\newcommand{\be}{\begin{equation}}
\newcommand{\ee}{\end{equation}}
\newcommand{\f}{\frac}
\newcommand{\s}{\sqrt}

\newcommand{\bea}{\begin{eqnarray}}
\newcommand{\eea}{\end{eqnarray}}
\newcommand{\ba}{\begin{align}}
\newcommand{\ea}{\end{align}}

\newcommand{\la}{\langle}
\newcommand{\ra}{\rangle}
\newcommand{\beq}{\begin{equation}}
\newcommand{\eeq}{\end{equation}}

\DeclareMathOperator{\tr}{tr}

\newcommand{\al}{\alpha}
\newcommand{\ve}{\varepsilon}

\title{
 Holographic R\'enyi relative divergence in  JT gravity
}

\author[a,b]{\! Tomonori Ugajin}

\affiliation[\,a]{Center for Gravitational Physics,
Yukawa Institute for Theoretical Physics, Kyoto University,
Kitashirakawa Oiwakecho, Sakyo-ku,
Kyoto 606-8502, Japan}
\affiliation[\,b]{The Hakubi Center for Advanced Research, Kyoto University,
Yoshida Ushinomiyacho, Sakyo-ku, Kyoto 606-8501, Japan}

\emailAdd{tomonori.ugajin@yukawa.kyoto-u.ac.jp}

\abstract{
We holographically  compute the R\'enyi relative divergence  $D_{\alpha} (\rho_{+} || \rho_{-})$ between two density matrices $\rho_{+}, \; \rho_{-}$  prepared by  path integrals with  constant background fields $\lambda_{\pm}$ coupled to  a marginal operator  in JT gravity.  Our calculation is non perturbative in the difference between two sources $ \lambda_{+} -\lambda_{-}$. When this difference is large, the bulk geometry becomes a black hole with the maximal temperature  allowed by the  R\'enyi index $\alpha$.  In this limit,  we find an analytic expression of the R\'enyi relative divergence, which is given by the on shell action of the back reacted black hole plus the contribution coming from the discontinuous change of the background field. 
}


\begin{document}

\maketitle

\section{Introduction}

%

\vspace{0.2cm} 

Distance measures on a space of density matrices play a crucial role in quantum information theory. Among them, relative entropy \cite{umegaki1962conditional} has nice properties such as positivity and monotonicity, which have been efficiently used to prove many intriguing results in quantum field theory, see for example\cite{Wall:2011hj,Bousso:2014sda,Faulkner:2016mzt}. The relative entropy $S(\rho|| \sigma)$ between two density matrices $\rho, \sigma$ is defined by 
\be 
S(\rho|| \sigma) ={\rm tr} \rho \log \rho - {\rm tr} \rho \log \sigma . 
\ee 

\vspace{0.2cm} 
Properties of relative entropy are reviewed in \cite{Witten:2018zxz}.

Although it is difficult to explicitly calculate relative entropy in QFT, we can perturvatively compute $S(\rho_{0}+\delta \rho|| \rho_{0}) $ by  expanding it in $\delta \rho$, when the two density matrices are sufficiently close to each other $ |\delta \rho| \ll 1$.
A nice way to do this is to use the modular flow of the reference state $\rho_{0}$
\cite{Faulkner:2014jva}.  When this prescription is applied to  conformal field theory, it turned out the second order term of the expansion (Fisher information) can be rewritten holographically in terms of the canonical energy on the dual anti de Sitter space \cite{Hollands:2012sf,Lashkari:2015hha}. The result matches with the prediction of the Ryu Takayanagi formula and its covariant generalization \cite{Ryu:2006ef, Ryu:2006bv, Hubeny:2007xt}. See the related discussions \cite{Sarosi:2016atx,Sarosi:2016oks,Sarosi:2017rsq,Faulkner:2017tkh,Lashkari:2018oke,Bhattacharya:2012mi, Blanco:2013joa,Wong:2013gua,Haehl:2017sot,Faulkner:2013ica}.

\vspace{0.2cm} 

R\'enyi relative divergence (RRD) is a one parameter generalization of relative entropy. It is defined by
\cite{PETZ198657}, 

\be 
D_{\alpha}(\rho|| \sigma) = -\f{1}{1-\al}\; \log {\rm tr}\; \left[\rho^{\al}\; \sigma^{1-\al} \right]. 
\ee 

It has also nice properties which are inherited from relative entropy \cite{DBLP:journals/corr/abs-1206-2459}. Recent discussions of applications of RRD  include \cite{2015PNAS..112.3275B, Casini:2018cxg,Moosa:2020jwt,deBoer:2020snb,Lashkari:2018nsl}.  One of the difficulties in computing RRD is that 
this quantity breaks the $U(1) $ translational symmetry of  Euclidean time direction. 
In \cite{Ugajin:2018rwd}, an efficient way to perturbatively expand RRD  by utilizing resolvent of  density matrix was found. In the same paper, it turned out that the second order term of the expansion of RRD again has a holographic expression.  Other studies of perturbative expansion of RRD include \cite{Bao:2019aol,May:2018tir,Bernamonti:2018vmw}. 

\vspace{0.2cm}

These perturbative expansions of relative entropy as well as RRD allow us to efficiently compute them. However it turned out that in general these series do not converge, 
as is often the case in perturbations in quantum field theory. This is because $\delta \rho$ is not a bounded operator. Therefore finding a prescription 
to resum these series is important. This is also interesting from the holographic point of view, as the higher order terms of the expansions are related to the emergence of full bulk gravitational equations of motion beyond the second order. See also  \cite{Balakrishnan:2020lbp} for a recent progress to resolve the non convergence problem. 

\vspace{0.2cm} 

As a first step toward this direction, in this paper we holographically compute  Renyi relative divergences without using  perturbative expansion,  in two dimensional JT gravity \cite{Jackiw:1984je,TEITELBOIM198341}  plus a bulk  matter theory. We hope this example will give a hint on how to resum the perturbative series found in our previous papers. 
JT gravity describes near horizon dynamics of near extremal black holes \cite{Almheiri:2014cka}. The dynamical degrees of freedom of the theory are the reparametrization modes, which live on the boundary of $AdS_{2}$, break 
the asymptotic symmetry of the spacetime \cite{Maldacena:2016upp}. The Lagrangian of the resulting effective theory on the boundary is given by the Schwarzian derivative of the reparametrization modes. A nice review of this topic is \cite{Sarosi:2017ykf}.

\vspace{0.2cm} 

We prepare two density matrices by turning on constant background fields coupled to a marginal operator operator in the boundary theory. We then calculate the RRD between them, which has a path integral expression with a time dependent background field. In order to calculate this RRD holographically, we first find the bulk configuration whose boundary condition matches with  the time dependent source  the field theory side. This bulk configuration, known as Janus solution was obtained in  \cite{Bak:2018txn}.  In this paper we will fully make use of this result \footnote{This solution shares  similar properties with its higher dimensional counter part \cite{Bak:2007jm, Bak:2007qw}.}. A  nice thing about the solution in two dimension is that the simplicity of the theory allows us to treat the effect of back reaction exactly for general R\'enyi index $\al$. In higher dimensions, what we can do at best is assuming the source is small, treat the back reaction of the source perturbatively and calculate the change of the on shell action. 

\vspace{0.2cm} 

This paper is organized as follows. 
In section \ref{section:setup} we define the R\'enyi relative divergence of our interest in Schwarzian theory  coupled to a marginal operator.   Also, we  explain the path integral expression of the RRD in this theory. In section \ref{section:bulk} we discuss the holographic dual of this setup, ie JT gravity in  $AdS_{2}$ coupled to a bulk massless scalar field.  We then present the relevant solution, whose on shell action computes the R\'enyi relative divergence holographically.  In section \ref{section:rep}, we derive the reparametrization mode of the boundary theory which dominates the path integral, from the bulk configuration derived in the previous section, and carefully evaluate its Schwarzian action in section \ref{section:action}. 
We then assemble these calculations  and present the main result in section \ref{section:RRD}.  In section \ref{section:Perturb}, we discuss the perturbative  expansion of the main result.

\section{The set up} 
\label{section:setup}

 We study a $0+1$ dimensional theory of  reparametrization modes $f(u) \in {\rm Diff}\;[S^{1}]/SL(2,\mathbb{R})$, on the circle $S^{1}$  with periodicity $2 \pi$. The action of this theory is given by 
 Schwarzian derivative of $f(u)$,  
 
 \be
 S_{{\rm Sch}}[f] =-C \int_{0}^{2\pi} du \; \left[\{ f(u),u \} +\f{f'^{2}}{2} \right], \quad \{ f(u),u \} = -\f{1}{2} \left(\f{f''}{f'} \right)^{2} + \left(\f{f''}{f'} \right)'. 
 \label{eq:Schaction}
\ee 

Throughout this paper,  $u$ is the coordinate of $S^{1}$ on which the theory is defined.  This action naturally appears as the low energy effective theory of the SYK model \cite{Maldacena:2016hyu}.

 We couple these modes  to an scalar operator $\mathcal{O}$. For simplicity we assume the operator  is marginal, ie, its conformal dimension  $\Delta$ is $1$.
 In this theory,  there is a class of density matrices $\{\rho_{\lambda}\}$,  prepared by turning on a constant background field $\lambda \in \mathbb{R}$  for the marginal operator $\mathcal{O}$  on  the thermal ensemble  $e^{-2\pi H}$, 
\be 
\rho_{\lambda} =\exp \left[-2\pi H +\lambda \int_{0}^{2\pi} d u \; \mathcal{O}(u) \right], 
\ee 
where $H$ is Hamiltonian of Schwarzian theory.  We  will take into account  the normalization of the density matrix later.

We would like to compute the R\'enyi relative divergence$D_{\al}(\rho_{\lambda_{+}}||\rho_{\lambda_{-}})$ between two such density matrices $\rho_{\lambda_{+}}$ and  $\rho_{\lambda_{-}}$. To this end, we first calculate  following quantity,
\begin{align} 
  {\rm tr} \;\rho_{\lambda{+}}^{\alpha}\; \rho_{\lambda_{-}}^{1-\alpha}=  {\rm tr} \left[ \exp \left( -H + \lambda_{+}\int^{2\pi \alpha}_{0} \mathcal{O} (u) \:\;du + \lambda_{-} \int^{2\pi}_{2\pi \alpha} \mathcal{O} (u) \;du \right) \right] \label{eq;relativeren}. 
\end{align}

The right hand side has a path integral expression, 
\be 
 {\rm tr} \;\rho_{\lambda{+}}^{\alpha}\; \rho_{\lambda_{-}}^{1-\alpha}= \int Df  \; \exp \left[  -S_{{\rm Sch}}[f] -\f{1}{8\pi}\int^{2\pi}_{0} du_{1} du_{2}\lambda(u_{1})\lambda(u_{2})\left(\f{ f'(u_{1}) f'(u_{2})}{\sin \left(\f{f(u_{1})-f(u_{2})}{2} \right)^{2}}\right) \right],
\ee

\begin{figure}[t]
    \centering
    \includegraphics[scale=.2]{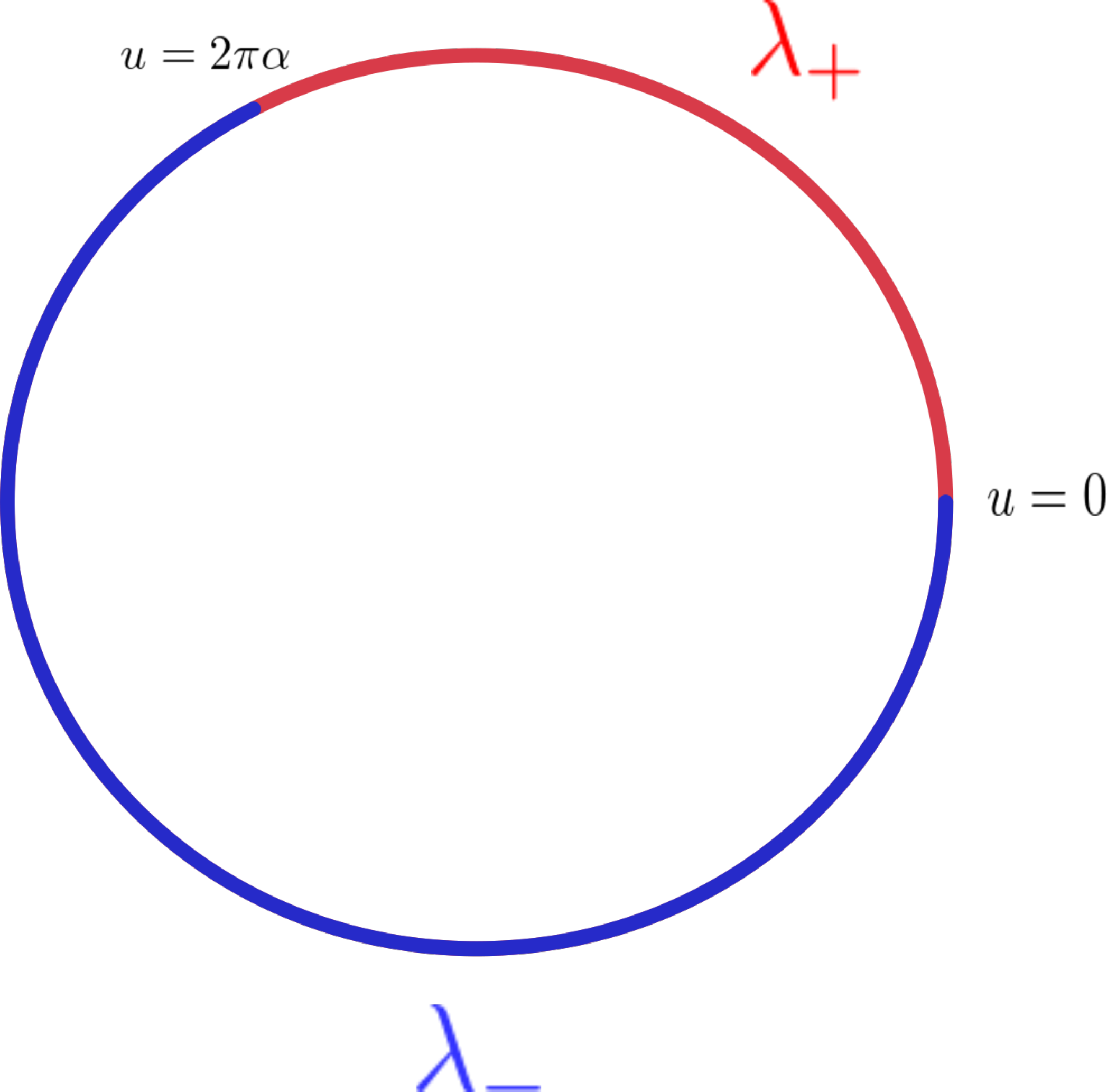}
    \caption{\small{The background field profile $\lambda(u)$  for the marginal operator which appears in the path integral.    The source  jumps discontinuously at  $ u=0$ and   $ u= 2\pi \alpha$ 
 }}
  \label{fig:S1}
\end{figure}

with the time dependent  background field $\lambda(u)$  which is coupled to the marginal scalar operator $ \mathcal{O}$, 

\be 
\lambda (u) = 
\begin{cases} 
\lambda_{+} & \quad 0<u<2\pi \alpha \\ 
\lambda_{-} & \quad 2\pi \alpha<u<2\pi,
\end{cases} 
\label{eq;bdycond} 
\ee 
see Fig \ref{fig:S1}.
We evaluate this path integral  in the semi classical limit  $C \gg 1$.

\section{ The holographic set up} \label{section:bulk}

The setup we  are considering has a gravity dual,  namely $AdS_{2}$ JT gravity coupled to a free massless scalar field $\chi$ which is dual to the marginal operator $\mathcal{O}$.   The total bulk  action is

\begin{equation} 
I=-\frac{\phi_{0}}{16\pi G}\left[ \int dx^{2} \s{g} R +2\int K \right] -\frac{1}{16\pi G}\left[ \int dx^{2} \s{g} \Phi(R +2) +2\int \phi_{b} K \right] +I_{M}[g, \chi] , \label{eq:gaction} 
\end{equation} 

JT gravity  part involves dilaton $\Phi$ as well as metric $g_{\mu\nu}$. The matter part 
$I_{M}[g, \chi]$ is  the action of free  scalar field $\chi$, 
\be 
I_{M}=-\f{1}{2} \int dx^{2} \s{g} \left( g^{ab}\nabla_{a}\chi \nabla_{b} \chi -m^2 \chi^2 \right). 
\ee 

One notable feature of the bulk action \label{eq; action} is that  there is no coupling between dilaton $\phi$ and the matter field $\chi$. Hereafter we set the mass of the scalar field to zero $m=0$, so that the primary $ \mathcal{O}$ is a marginal operator $\Delta_{\mathcal{O}} =1$.

This action describes excitations on top of an  higher dimensional extremal black hole geometry, dimensionally reduced to two dimensions. 
The first term of the action \eqref{eq:gaction}  is purely topological, describes the grand state degeneracy of the extremal black hole. Therefore, the constant  $\phi_{0}$ corresponds to the entropy of the higher dimensional extremal black hole. The second term of the action \eqref{eq:gaction}  captures excitations on top of the extremal black hole, and the total dilaton  $\Phi +\phi_{0}$  corresponds to  the entropy  of the  excited black hole. In order for the effective description valid,  we need  $\phi_{0} \gg \Phi$. The equations of motion for dilaton $\Phi$ and the metric $g_{\mu\nu}$ is 

\be 
\nabla_{\mu} \nabla_{\nu}\Phi-g_{\mu\nu} \nabla^{2} \Phi- g_{\mu\nu} \Phi =8\pi G\; T_{\mu\nu}[\chi],\quad  \quad R+2 =0, \label{eq:eomforrrd} 
\ee 

where  the stress tensor $T_{\mu\nu}[\chi]$ of the massless scalar field $\chi$ ,
\be 
T_{\mu\nu}[\chi]=\nabla_{\mu}\chi \nabla_{\nu} \chi -\f{g_{\mu\nu}}{2} \left( \nabla^{\rho}\chi \nabla_{\rho}\chi \right). 
\ee 

The second equation of  \eqref{eq:eomforrrd} fixes the metric to that of pure $AdS_{2}$. In Poincare coordinates, this metric is given by 
\be 
ds^2 = \f{dt^2 +dz^2}{z^2} \label{eq;poin} 
\ee

Due to the equations of motion \eqref{eq:eomforrrd}, the bulk action \eqref{eq:gaction} is  reduced to  the field theory  action \eqref{eq;relativeren} on $S^{1}$ \cite{Maldacena:2016upp}, once we fix the location of the boundary $(t, z)=(t(u), z(u))$ in the bulk, and the relation between $t(u)$ and the reparametrization mode $f(u)$, which will be discussed later. In doing so, we also need the identification of  the coupling constants on both sides, 
\be 
C=\f{\bar{ \phi}}{8\pi G}.
\ee

$t(u)$ and $z(u)$  define an embeddins of the boundary $S^{1}$ to Poincare $AdS_{2}$. Furthermore,  symmetry imposes the relation $z(u)=\ve t'(u)$,  where $\ve$ is the UV cut off of the boundary theory.
 The location of the boundary line  $t(u)$ is specified by imposing the boundary conditions 
\be 
\Phi (t(u), \varepsilon t'(u)) = \f{\bar{ \phi}}{\varepsilon}, \quad  \chi (t(u), \ve t'(u))= \lambda (u). \label{eq;bdydil} 
\ee 

 In this way, the task of computing  R\'enyi relative divergence in the semi classical limit $C \gg 1$   is reduced  to find  the corresponding solution of the gravitational equations motion \eqref{eq:eomforrrd}, and specifying the reparametrization mode $f(u)$ from  the  solution.

\subsection{The construction of the bulk configuration.} 

\subsubsection{ The Janus solution} 

Our starting point is the Janus solution, obtained in \cite{Bak:2018txn}. This solution also naturally appears in a recent discussion on  entanglement entropy between two disjoint universes and its relation to island formula \cite{Balasubramanian:2020coy}. Also, this class of solutions has been also discussed in slightly different contexts \cite{Chen:2020tes,Garcia-Garcia:2020ttf}. In the global AdS  coordinates $(\mu, t_{s})$   where the metric takes following form,
\be 
ds^{2} = \f{dt_{s}^2 +d\mu^2}{\cos^2 \mu},\quad -\f{\pi}{2}<\mu < \f{\pi}{2},
\ee

 this solution  is given by 

\be 
\Phi (\mu, t_{s})= \Phi_{0}(\mu, t_{s}) -4\pi G\gamma^2 (1+ \mu \tan \mu), \label{eq:dilsol}
\ee 

and 
\be 
\chi (\mu) =\gamma \left(\mu-\f{\pi}{2} \right) + \lambda _{+}, 
\ee

where $\gamma$ is  the difference between the background fields, $\gamma = (\lambda_{-}- \lambda_{+})/\pi$, which is necessary to impose the boundary condition, 
\be 
\chi\big|_{\mu=\f{\pi}{2}} =\lambda_{+}, \quad \chi\big|_{\mu=-\f{\pi}{2}} =\lambda_{-}
\ee

$\Phi_{0}(\mu, t_{s}) $ is the "souceless" of the dilaton, satisfying the equations of motion  \eqref{eq:eomforrrd} with $T_{\mu\nu}[\chi] =0$. This part will be fixed later.

\subsubsection{Mapping the solution to Poincare AdS} 

In order to relate the dilaton profile \eqref{eq:dilsol} to the reparametrization mode $f(u)$, it is convenient to  write down the solution in Poincare coordinates
 with the metric (\ref{eq;poin}).

The relation between global and Poincare coordinates is 
\be 
z+it =e^{i\mu +t_{s}} \quad \leftrightarrow \quad t_{s} =\f{1}{2} \log \left[ z^2+t^2 \right], \quad \mu = \tan^{-1} \f{t}{z}. \label{eq;coordrel} 
\ee 
The dilaton profle in Poincare coordinates is,

\be 
\Phi (z,t)=\Phi_{0}(z,t) - 4\pi G \gamma^{2}\left( 1+ \f{t}{z} \tan^{-1}\f{t}{z} \right). \label{eq;dilpoincare} 
\ee 

 the sourceless part of dilaton $\Phi_{0}$  is in general,
\be 
\Phi_{0}(z,t) = \f{\xi(t^2+z^2)+\kappa t +\eta }{z} .\label{eq:nodefdi} 
\ee

Finally, the scalar field $\chi$ profile is 
\be 
\chi(t,z) = \gamma \left( \tan^{-1} \f{t}{z} -\f{\pi}{2} \right) +\lambda_{+}. \label{eq;scalpoinc} 
\ee 

Its boundary value depends on the sign of the time like coordinate $t$,
\be 
\chi(t,z) \rightarrow 
\begin{cases} 
\lambda_{+} & t>0 \\ 
\lambda_{-} & t<0 
\end{cases} 
\quad ,z \rightarrow 0 
\ee

\section{ The on shell reparametrization mode} 

\label{section:rep}

Having specified the solution of the equations of motion,  \eqref{eq;dilpoincare}  and \eqref{eq;scalpoinc}, now the task is to find the location of the boundary $(t,z)=(t(u), \ve t'(u))$ on which the boundary conditions \eqref{eq;bdydil} are
satisfied. This also fixes the parameters $(\xi,\kappa,\eta)$ of the  sourceless part of dilaton $\Phi_{0}$ \eqref{eq:nodefdi}. At the same time we relate the boundary location $t(u)$ in Poincare AdS$_{2}$ to the boundary location in  Poincare disc, which is identified with the reparametrization mode $f(u)$  in the Schwarzian action \eqref{eq:Schaction},     via an appropriate bulk diffeomorphism.  

-

\subsubsection*{The boundary location for sourceless dilaton} 

We begin our discussion from the  case where dilaton is sourceless $\Phi =\Phi_{0}$, ie when the stress tensor of the scalar field vanishes  $T_{\mu\nu} [\chi]=0$. The general form of such dilaton profile is given by \eqref{eq:nodefdi}.  We want an embedding $t(u)$, which maps the segment $I^{+}: 0<u<2\pi \alpha $ (the red interval in Fig \ref{fig:S1}) on the boundary circle of Poincare disc $S^{1}$,  to $t>0$ part of the boundary of Poincare AdS$_{2}$,  and its complement  $I^{-}: 2\pi \alpha <u<  2\pi$ (the blue interval in Fig \ref{fig:S1}) to  $t <0$. Such a map is given by

\be 
t(u) =- \f{\sin \f{u}{2}}{ \sin \left(\f{u-2\pi \alpha}{2} \right)} \label{eq;bdal}.
\ee 
The sourceless dilaton profile which satisfies the boundary condtion \eqref{eq;bdydil}, is given by choosing  the coefficients $(\xi, \kappa, \eta)$ in \eqref{eq:nodefdi}, 
\be 
\xi=\f{\bar{\phi}}{\sin \pi \alpha} , \quad \kappa =\f{2\bar{\phi}}{\tan \pi \alpha} , \quad \eta =\f{\bar{\phi}}{\sin \pi \alpha}. \label{eq;coffs} 
\ee 

Note that  since the Schwarzian derivative of (\ref{eq;bdal}) is still constant 
\be 
\{ t(u),u\}=\f{1}{2},
\ee 

 the introduction of the index $\al$ does not change the thermodynamic quantities of the black hole as it should be.

\subsubsection*{The  boundary location in the presence of the source}

Now let us turn on the source $\gamma$. We find that by choosing following coefficients $(\xi, \kappa, \eta)$ in \eqref{eq:nodefdi}, one can satisfy the boundary condition.

\be 
\xi =\f{\bar{\phi} \nu }{\sin \pi \al \nu}, \quad \kappa = \f{2\bar{\phi}\nu}{\tan \pi \al \nu} +2\pi^2 G \gamma^2, \quad \eta = \f{\bar{\phi}\nu}{\sin \pi \al \nu} \label{eq;coff1} .
\ee 

Here we introduced $\nu$, which will be identified with the temperature of the backreacted  black hole,  caused by the source $\gamma$.

 We should remark that  our choice of sourceless part $\Phi_{0}$ is slightly different from the choice of \cite{Bak:2018txn}. In \cite{Bak:2018txn},  $\Phi_{0}$ was fixed by demanding that the temperature $\nu$ of the back reacted  black hole is always fixed, ie independent  of $\gamma$.   This is not convenient for our purpose, since  with this choice the boundary conditions \eqref{eq;bdydil}  can not be satisfied. 
Indeed,   we will see  later that onece we choose  $\Phi_{0}$ which satisfies the boundary condition,  then  the temperature of the  backreacted black hole does depend on $\gamma$.

\vspace{0.3cm}
Let us first consider the interval $0< u<2\pi \alpha$. If we choose the map,
\be
t(u)=-\f{\sin \f{\nu u}{2}}{\sin\left( \f{\nu u -2\pi \alpha \nu }{2} \right)}
\ee
then $I^{+}$ is correctly  mapped to $t>0$, and satisfies the boundary condition for the scalar field $\chi(u)=\lambda_{+}$.  Furthermore, due to the choice of the sourceless part,the pull back of  total dilaton on $I^{+}$ also satisfies the boundary condition  $\Phi (t(u), \ve t'(u))=\bar{\phi}/\ve$.

On the complement $I^{-}: 2\pi \alpha < u<2\pi$,  we  employ  a generalized ansatz for $t(u)$,
\be 
t(u) =- \f{\sin \f{f(u)}{2}}{ \sin \left(\f{f(u)-2\pi \nu \alpha }{2} \right)} \label{eq;bddel}.
\ee 
Also, for the continuity we demand $f(2\pi \al) =2\pi \al \nu$.
Then the pullback of the dilaton profile \eqref{eq;dilpoincare} with \eqref{eq;coff1} on this curve is

\be 
\Phi (u) \equiv \ve \Phi (t(u), \ve t'(u))= \f{\phi_{0}}{f'} -4\pi G \gamma^{2} \left[ -\f{\pi}{2} +\tan^{-1} \left( \f{h(u)}{\varepsilon f'} \right) \right] \left( \f{h(u)}{\varepsilon f'} \right) , \label{eq:pulldil}
\ee 

with 
\be 
h(u) = \f{-2 }{\sin (\pi \nu \alpha) }\sin \f{f}{2}\sin \left(\f{f-2\pi \nu\alpha}{2} \right). \label{eq;hu} 
\ee 

When $|h(u)| \gg \varepsilon f'(u)$, we can replace arectan in \eqref{eq:pulldil} with either $\f{\pi}{2}$ or $-\f{\pi}{2}$, depending on the sign of $h(u)$. For example we can do this on the regions 
$I^{+}_{\ve}: \s{\ve} < u< 2\pi \alpha -\s{\varepsilon}$ and $ I^{-}_{\ve} :2\pi \alpha +\s{\ve} <u< 2\pi -\s{\ve}$, where  $h(u)$ is sufficiently large,
in the $\ve \rightarrow 0$ limit.

If we impose the boundary condition in \eqref{eq;bdydil} for the dilaton,  we get
\be 
f'(u) = 
\begin{cases} 
\nu, \quad u \in I^{+}_{\ve}, \\[+10pt] 
\nu - \f{8\pi^{2} \gamma^{2} G}{\bar{\phi}\sin (\pi \nu \alpha)}\sin \f{f}{2}\sin \left(\f{f-2\pi \nu \alpha}{2} \right) \quad u \in I^{-}_{\ve} . 
\end{cases} 
\label{eq:approxf} 
\ee 

\hspace{0.3cm}

\subsection{Fixing the temperature }

Now let us fix the temperature $\nu$ of the backreacted black hole.  In order to do this, recall the expression  the two point function of the marginal operator $\mathcal{O}$ dual to the bulk scalar field $\chi$, 

\be 
\la \mathcal{O} (u_{1}) \mathcal{O} (u_{2}) \ra = \f{f'(u_{1}) f'(u_{2})}{4\sin^{2} \left(\f{f(u_{1})- f(u_{2})}{2} \right)}.
\ee

In order for the correlator 
to have the appropriate periodicity $u_{1} \sim u_{1}+2\pi$, the reparametrization mode $f(u)$ has to satisfy $f(2\pi)= 2\pi$. From \eqref{eq:approxf}, 
we see this condition is equivalent to

\be 
\int^{2\pi}_{2 \pi \alpha \nu} \f{df}{ \left[\nu - \f{8\pi^{2} \gamma^{2} G}{\bar{\phi}\sin (\pi \nu \alpha)}\sin \f{f}{2}\sin \left(\f{f-2\pi \nu \alpha}{2} \right) \right]} =2\pi (1-\alpha). \label{eq:tempeq}
\ee 

This equation determines the temperature $\nu $ of the back reacted black hole, as a function of $\gamma$ and $\al$. In general, it is difficult to obtain its explicit solution.However,
one can perturbatively solve this equation around $\gamma =0$, let $\nu =1+\delta \nu$, then, 
\be 
\delta \nu =\f{2\pi \gamma^{2} G}{\bar{\phi}\sin (\pi \alpha)} \left[2\pi (1-\alpha) \cos \pi \alpha + 2\sin \pi \alpha \right] + O (\gamma^{4}) .  \label{eq:perturb}
\ee

\subsection{The large source limit  } 
When the difference of the source $\gamma$ is large, we can also solve this equation  exactly.  In this limit,
 the temperature of the back reacted black hole  gets maximal, ie  $\nu \rightarrow 1/\alpha$. If the temperature could be larger than this value,  we were unable to impose the  periodicity condition $f(2\pi ) =2\pi$. Indeed,  when $\gamma \gg 1$ one can consistently solve the equation  for $\nu$ by $1/\gamma^{2}$ expansion. 

Since  $ \nu \rightarrow 1/\alpha $, the range of the integration \eqref{eq:tempeq},  become smaller, $2\pi \alpha \nu \rightarrow 2\pi$.  In order for the integral non zero, the integrand has to diverge in this limit. In particular, by expanding the integrand near  $f\sim  2\pi$, we see that in order for this to happen, $\nu$ has to satisfy,
\be 
\nu = \f{1}{\alpha} -\f{\bar{\phi}}{2\alpha \pi^{3} G \gamma^{2}}  +\cdots  .
\label{eq:templarge}
\ee

We emphasize that this reasoning is only  valid  when $\alpha \neq 1$ or equivalently, when the left hand side of \eqref{eq:tempeq} is non zero.   In particular, when we calculate the relative entropy, 
\be
S(\rho_{+}|| \rho_{-}) = \lim_{\alpha \rightarrow 1}  D_{\alpha}(\rho_{+}|| \rho_{-}),
\ee
we can not use the expansion, since this limit involves  $\alpha \sim 1$ .  

\section{ Evaluating the semi classical action} 
\label{section:action}

Having specified the reparametrization mode which dominates the path integral \eqref{eq;relativeren}, in this section, we evaluate its on shell action. 
It is straightfoward to  evaluate the action on $I^{\pm}_{\ve}$ where the reparametrization mode  is given by \eqref{eq:approxf}. However near $ u=0 \; 2\pi \alpha$,  lets say $C^{+}_{\ve}: -\s{\ve}< u< +\s{\ve} $, and  $C^{-}_{\ve}: 2\pi \al -\s{\ve}< u<2\pi \al +\s{\ve}$,    the boundary source $\lambda (u)$ changes discontinuously, and total dilaton profile behaves like a step function. Accordingly, higher order derivatives of the reparametrization mode develop delta functional singularities.   Therefore, in evaluating  its Schwarzian action $S_{{\rm Sch}}$, we need a special care.  It is convenient to split the integral into four pieces. 
\begin{align} 
S_{{\rm Sch}} &=-C\; \lim_{\ve \rightarrow 0}\left[\int_{C^{+}_{\ve}} +\int_{I^{+}_{\ve}} + \int_{C^{-}_{\ve}} +\int_{I^{-}_{\ve}} \right] du \left[\f{f'^{2}}{2} -\f{1}{2}\left(\f{f''}{f'} \right)^{2}+\left(\f{f''}{f'} \right)' \right] \label{eq:Lagrangian}\\[+10pt]
&\equiv \lim_{\ve \rightarrow 0} \;  S_{C^{+}_{\ve}} +S_{I^{+}_{\ve}}+S_{C^{-}_{\ve}}+S_{I^{-}_{\ve}}
.
\end{align}

\subsection{Normal Part}

It is straightfoward to evaluate $S_{I^{+}_{\ve}}$ and 
$S_{I^{-}_{\ve}}$. This is because the expression of the reparametrization mode gets simplified in these regions \eqref{eq:approxf}. In general, when a reparametrization mode satisfies the relation,
\be 
f' = a-b \cos (f-\delta), \quad \delta \in \mathbb{R},
\ee
then, the Lagrangian is constant, 
\be 
L_{{\rm Sch}}[f] =\f{f'^{2}}{2} -\f{1}{2}\left(\f{f''}{f'} \right)^{2}+\left(\f{f''}{f'} \right)' = \f{a^{2}-b^{2}}{2}.
\ee

For the remparametrization mode of interest \eqref{eq:approxf}, we obtain

\be
L_{{\rm Sch}}[f]=
\begin{cases} 
\f{\nu^{2}}{2},\quad  u \in I^{+}_{\ve} \\[+10pt] 
\f{1}{2}\left(\nu- \f{4\pi^{2} G \gamma^{2}}{\bar{\phi} \tan \pi \alpha \nu} \right)^{2} -\f{1}{2}\left( \f{4\pi^{2} G \gamma^{2}}{\bar{\phi} \sin \pi \alpha \nu} \right)^{2},\quad u \in I^{-}_{\ve}
\end{cases} 
\label{eq:lsch}
\ee

$L_{{\rm Sch}} $ is related to the stress energy expectation value $\la T_{00} \ra$ of the boundary theory \cite{Maldacena:2016upp}
\be 
\la  T_{00} \ra = C L_{Sch}[f].
\ee

Therefore the expression \eqref{eq:lsch} implies, when the dilaton profile  is continued to Lorenzian signature, it represents a two sided black hole, in which temperatures of left and right horizons are  different. However  in the large $\gamma$ limit, we can see  that
\be 
\f{1}{2}\left(\nu- \f{4\pi^{2} G \gamma^{2}}{\bar{\phi} \tan \pi \alpha \nu} \right)^{2} -\f{1}{2}\left( \f{4\pi^{2} G \gamma^{2}}{\bar{\phi} \sin \pi \alpha \nu} \right)^{2}  \rightarrow \f{\nu^{2}}{2},
\label{eq:tempcoincide}
\ee
by plugging \eqref{eq:templarge}.
This indicates that two black hole temperatures coincide in this limit.

By combining these results, the normal part of the action is given by 
\be
\lim_{\ve \rightarrow 0} \;   S_{I^{+}_{\ve}}+S_{I^{-}_{\ve}}
=-  \f{\bar{\phi}}{8\pi G} \left\{ \pi \alpha \nu^{2} + \pi (1-\alpha)\left[ \left(\nu- \f{4\pi^{2} G \gamma^{2}}{\bar{\phi} \tan \pi \alpha \nu} \right)^{2} -\left( \f{4\pi^{2} G \gamma^{2}}{\bar{\phi} \sin \pi \alpha \nu} \right)^{2} \right] \right\}.
\ee

\subsection{Defect part}

Now let us evaluate the action on the transition regions $C^{+}_{\ve}$, $C^{-}_{\ve}$.  Naively speaking, since we take $\ve \rightarrow 0$,  $S_{C^{+}_{\ve}}$  $S_{C^{-}_{\ve}}$ are  vanishing. However, we see that  since the last term of the Lagrangian \eqref{eq:Lagrangian} is highly singular due to the discontinuity of the source profile $\lambda(u)$, and $S_{C^{+}_{\ve}}$  $S_{C^{-}_{\ve}}$ are indeed non vanishing even in the $\ve \rightarrow 0$ limit. Indeed the sum of these two terms  are given by 
\begin{align} 
\lim_{\ve \rightarrow 0} S_{C^{+}_{\ve}}+S_{C^{-}_{\ve}}& =-C \lim_{\ve \rightarrow 0} \left[\left(\f{f''}{f'}\right)_{u=\s{\ve}} -\left(\f{f''}{f'}\right) _{2\pi -\s{\ve}}  \right]
-C \lim_{\ve \rightarrow 0} \left[\left(\f{f''}{f'}\right)_{u=\s{\ve}} -\left(\f{f''}{f'}\right) _{2\pi -\s{\ve}}  \right] \nonumber\\[+10pt]
&=-\pi \gamma^{2}.
\label{eq:discaction}
\end{align}

Here, we  used the fact that the integral of the first two terms of the Lagrangian vanishes in the $\ve \rightarrow 0$ limit.  Two terms of the right hand side of above equation can be evaluated by using the approximate expressions \eqref{eq:approxf} for $f(u)$.

\subsection{Matter  part} 

Finally, we need to evaluate the matter action related  to the marginal operator $\mathcal{O}$, which explicitly depends on the 
the time dependent source $\lambda(u)$ \eqref{eq;bdycond}.
\be 
-I_{M} =\f{1}{8\pi} \int^{2\pi}_{0} du_{1} \int^{2\pi}_{0} du_{2} \; \lambda(u_{1})\lambda(u_{2})\;  \f{f'(u_{1})f'(u_{2})} {\sin^{2} \left( \f{f(u_{1})-f(u_{2})}{2}\right)} .
\ee

In evaluating this integral, we have to properly regulate the   singularity at  $u_{1}=u_{2}$.  This regularization can be implemented  by  introducing a UV cutoff $\ve$ to the integral  as,
\begin{align}
-I_{M} &=\f{1}{8\pi} \int^{2\pi-\ve}_{\ve} du_{1} \left[\int^{2\pi}_{u_{1}+\ve} +  \int^{u_{1}-\ve}_{0} \right]du_{2} \; \lambda(u_{1})\lambda(u_{2})\;  \f{f'(u_{1})f'(u_{2})} {\sin^{2} \left( \f{f(u_{1})-f(u_{2})}{2}\right)} \nonumber\\[+10pt] 
&=\f{2 \lambda_{+}^{2}\alpha}{\ve} +\f{2 \lambda_{-}^{2}(1-\alpha)}{\ve}
-\pi\gamma^{2}\log \left|  \f{\sin \pi \al \nu}{\ve} \right|,
\label{eq:Matteraction}
\end{align} 

here, we used the definition $\gamma=(\lambda_{-}-\lambda_{+})/2\pi$.

\subsection{The net result}

By combining these results, 
The total action is given by 

\begin{align}
\log \;\rho_{+}^{\alpha}\rho_{-}^{1-\alpha} &=
-\lim_{\ve \rightarrow 0} \left[S_{I^{+}_{\ve}}+S_{I^{-}_{\ve}}+S_{C^{+}_{\ve}}+S_{C^{-}_{\ve}} \right]- I_{M} \nonumber \\[+10pt]
&=\f{\bar{\phi}}{8\pi G} \left\{ \pi \alpha \nu^{2} + \pi (1-\alpha)\left[ \left(\nu- \f{4\pi^{2} G \gamma^{2}}{\bar{\phi} \tan \pi \alpha \nu} \right)^{2} -\left( \f{4\pi^{2} G \gamma^{2}}{\bar{\phi} \sin \pi \alpha \nu} \right)^{2} \right] \right\} +\pi \gamma^{2} \nonumber\\[+10pt]
&+\f{2 \lambda_{+}^{2}\alpha}{\ve} +\f{2 \lambda_{-}^{2}(1-\alpha)}{\ve}
-\pi\gamma^{2}\log \left|\f{\sin \pi \al \nu}{\ve}\right|.\label{eq:finalaction}
\end{align}

\section{ The calculation of the R\'enyi relative divergence } 
\label{section:RRD}

Now  let us assemble the results so far,  to calculate the R\'enyi relative divergence  $D_{\alpha} (\rho_{+} || \rho_{-})$ of our interest. Including the normalization, this quantity is defined by 
\be 
D_{\alpha} (\rho_{+} || \rho_{-}) = -\f{1}{1-\alpha}\log \left[\f{{\rm tr} \left[\rho_{+}^{{\alpha}} \rho_{-}^{{1-\alpha}} \right]}{ {(\rm tr} \left[\rho_{+} \right] )^{\alpha} ({\rm tr} \left[\rho_{-} \right] )^{1-\alpha}} \right].
\label{eq:defrrd}
\ee

The denominator is  computed by  evaluating the on shell action in the presence of  constant  scalar field $\chi  = \lambda_{\pm}$,

\be
\log \tr \rho_{+}= \f{\bar{\phi}}{8G} +\f{2 \lambda_{+}^{2}}{\ve},\; \quad 
\log \tr \rho_{-}= \f{\bar{\phi}}{8G} +\f{2 \lambda_{-}^{2}}{\ve}.
\ee

By combining this with \eqref{eq:finalaction}, we arrive 
\begin{align}
D_{\alpha} (\rho_{+} || \rho_{-})&=-\f{\bar{\phi}}{8 G(1-\alpha)} \left\{  \alpha \nu^{2} +  (1-\alpha)\left[ \left(\nu- \f{4\pi^{2} G \gamma^{2}}{\bar{\phi} \tan \pi \alpha \nu} \right)^{2} -\left( \f{4\pi^{2} G \gamma^{2}}{\bar{\phi} \sin \pi \alpha \nu} \right)^{2} \right]-1 \right\}\nonumber \\[+10pt]
&+ \f{\pi \gamma^{2}}{1-\alpha} \left(-1+\log \left| \f{\sin \pi \al \nu}{\ve} \right|  \right)
\label{eq:mainresult}
\end{align}

The result involves the temperature of the back reacted black hole $\nu$, which is specified  by solving  \eqref{eq:tempeq}. In particular, when $\gamma$ is large, it is given by  \eqref{eq:templarge}.  From this we obtain an analytic expression of the R\'enyi relative divergence in this limit, 
\be 
D_{\alpha} (\rho_{+} || \rho_{-}) =-\f{\bar{\phi}}{8G \alpha^{2}} (1+\alpha) +\f{\pi \gamma^{2}}{1-\alpha} \left(-1+\log 
\left| \f{1}{\ve}\sin \left( \f{\bar{\phi}}{2\pi^{2} G \gamma^{2}}\right) \right| \right),
\label{eq:finalexpression}
\ee

In this expression, $\gamma^{2} \ll 1/\ve$ is understood, in order for the validity of the bulk effective theory.
We  remark that the approximate expression of the black hole temperature \eqref{eq:templarge} is not valid near $\alpha =1$. This is the reason why  $\alpha \rightarrow 1$ limit of  \eqref{eq:finalexpression} is not smooth.

\section{A perturbative expansion of the R\'enyi relative divergence} 
\label{section:Perturb}

So far, we  computed  the R\'enyi relative divergence  only in the large $\gamma$ limit. This is because we can solve  the equation for  the temperature of the backreacted black hole $\nu$ only in this limit.  Of course,  we can study  this divergence  in another limit ,   namely in small  $\gamma $ limit, by  the perturbative expansion  from $\gamma=0$.  

The non trivial part of the reparametrization mode can be obtained from 
\be 
\int^{f(u)}_{2 \pi \alpha \lambda} \f{df}{ \left[\nu - \f{8\pi^{2} \gamma^{2} G}{\bar{\phi}\sin (\pi \lambda \alpha)}\sin \f{f}{2}\sin \left(\f{f-2\pi \lambda \alpha}{2} \right) \right]} =u-2\pi \al. 
\label{eq:thecondition}
\ee 

One can perturbatively solve this equation near $\gamma=0$, 
\be 
f(u) =u+ \gamma^2 g(u) +O(\gamma^{4}). 
\ee 

Accordingly the on shell Schwarzian action can be expanded like 
\be 
S_{{\rm Sch}}[f] =\sum_{n=0}^{\infty} \gamma^{2n} S_{2n} =S_{0} +\gamma^{2} S_{2} +\gamma^{4} S_{4} +O(\gamma^{6}). 
\ee 

The leading correction $S_{2}$ is vanishing because of periodicity $g(0) =g(2\pi)$. Thus, the first non trivial part of the action starts from   
 $S_{4}$. This gives the leading gravitational contribution to the R\'enyi relative divergence. 

\eqref{eq:thecondition} is equivalent to

\be 
f'(u) =\nu -\f{8\pi^{2}\gamma^{2} G}{\bar{\phi} \sin \pi \al} \sin\f{u}{2}\sin\f{u -2\pi \al}{2} +O(\gamma^{4}), \quad u \in I^{-}_{\ve}. 
\ee

By integrating this expression we get, 
\be 
g(u)= \left[ \delta \nu -\f{4\pi^{2}\gamma^{2} G}{\bar{\phi} \sin \pi \al} \left( (u-2\pi \al \right)\cos \pi \al - \left( \sin (u-\pi \al) -\sin \pi \al \right)\right], \label{eq;gu} 
\ee 
where $\delta \nu$ is given by \eqref{eq:perturb} which we reproduce here, 
\be 
\delta \nu = \f{2\pi \gamma^{2}G }{\bar{\phi}}\left[ 2\pi (1-\al) \cot \pi \al +2\right].
\ee

Notice that $g(u)$ satisfies the periodicity condition  $g(u)=g(2\pi)=0$. One can easily expand the Schwarzian action, up to $\gamma^{4}$ term, 
\begin{align} 
\gamma^{4}S_{4} &= \int^{2\pi}_{0} du \; (g'(u) )^{2}-(g''(u) )^{2} \\ \nonumber 
&= \int^{2\pi \al}_{0} du (g'(u) )^{2} +\int^{2\pi \al}_{0} du \; (g'(u) )^{2}-(g''(u) )^{2}.
\end{align} 

By plugging (\ref{eq;gu}) into the action, we obtain 
\be 
\gamma^{4}S_{4} =-2\pi ( \delta \nu)^{2} +\f{8\pi^{3}\gamma^{2} G}{\bar{\phi}} \; \delta \nu \;\cot \pi \al.
\ee 

Similarly, we can expand the matter part \eqref{eq:Matteraction} up to $\gamma^{4}$ term,

\be 
-I_{M} = \f{2\lambda_{+}^{2}}{\ve} - \pi \gamma^{2} \left[\log \left|  \f{\sin \pi \al}{\ve} \right| +\pi \al \cot \pi \al \; \delta \nu \right].
\ee

By assembling these results, we obtain the perturbative expansion of the R\'enyi relative divergence  \eqref{eq:defrrd} up to $\gamma^{4}$ order, 
\begin{align}
D_{\al} (\rho_{+}|| \rho_{-}) =-\f{1}{1-\alpha}\left(-2\pi ( \delta \nu)^{2} +\f{8\pi^{3}\gamma^{2} G}{\bar{\phi}} \; \delta \nu \;\cot \pi \al \right)+\f{\pi\gamma^{2}}{(1-\alpha)} \left[\log \left|  \f{\sin \pi \al }{\ve} \right| +\pi \al \cot \pi \al \; \delta \nu \right].
\end{align}

The relative entropy $S(\rho_{+} || \rho_{-})$ is given by  the $\alpha \rightarrow 1$ limit of the divergence, 

\be
S(\rho_{+} || \rho_{-}) ={\rm tr} \; \rho \log \rho -{\rm tr} \; \rho \log \sigma
= \lim_{\al \rightarrow 1} D_{\al} (\rho_{+}|| \rho_{-}).
\ee

This relative entropy has an additional divergence coming from the derivative of $\log| \sin \pi \alpha|$. After introducing the UV cut off for this, the perturbative expansion of the relative entropy is

\be 
S(\rho_{+} || \rho_{-})=\f{\gamma^{2}}{\ve} -\f{4\pi^{4}G \gamma^{4}}{3\bar{\phi}}-\f{2\pi}{3} \left( \f{4\pi^{2}\gamma^{2} G}{\bar{\phi} }\right)^{2}+O(\gamma^{4}) 
\ee

Positivity of the relative entropy  is guaranteed,  since the first term is positively divergent.

\section{Conclusion and discussions}

We studied  the R\'enyi relative divergence (RRD) $D_{\alpha} (\rho_{+} || \rho_{-})$ between  two states prepared by a path integral in the presence of constant back ground fields coupled to a marginal operator $\mathcal{O} $in $0+1$ dimensional Schwarzian theory. In doing so, we utilized the holographic set up, where JT gravity is coupled to a massless scalar field, dual to the marginal operator $\mathcal{O}$.

\vspace{0.2cm} 

The technical challenge of such a calculation is to  deal with the back reaction of the bulk  scalar field  which is sourced by the time dependent  background  field on the boundary.   Due to this  difficulty, the main tool to study  this kind of RRD has been the perturbative expansion with respect to the background field. However, the simplicity of JT gravity allows us to fully specify such  back reaction in an exact manner. Having this advantage in mind, in the body of this paper, 
we mainly focused on the regime where the back reaction of the source is significant. The back reaction made the  temperature $\nu$ of the black hole maximal, allowed by the R\'enyi  index $\alpha$, ie $\nu \rightarrow \f{1}{\alpha} $.  The resulting RRD was given by the on shell action of this black hole plus the contribution of the discontinuity of the source.

\vspace{0.2cm} 
 In \cite{Ugajin:2018rwd}, we obtained  a general formula  for  a perturbative expansion of R\'enyi relative divergence.   This was done by writing 
\be 
{\rm  tr }  \; \rho^{\alpha} \sigma^{1-\alpha} = \f{1}{2\pi i}\int_{C} dz \; z^{\alpha}  {\rm tr} \left[ \f{\sigma^{1-\alpha}}{z- \rho}\right],
\ee
and  expanding the the denominator of the right hand side, by assuming $\delta \rho= \rho-\sigma$ is small. 
The  result is concisely summarized  in terms of  integrals along the modular flow of the reference state $\sigma$.
Now, given the exact result \eqref{eq:mainresult} at hand, it is interesting to   check how the result in this paper is consistent with the formula for the perturbative expansion. In particular, we obtained $\gamma^{4}$ term of  the RRD. This should coincide with the result derived from the expansion formula.
Since the expansion obtained in this way is an asymptotic series, it would be interesting  to find a way to resum the series.  We believe our result gives a hint for this. 

\vspace{0.2cm} 

 Finally, let us briefly discuss the Lorenzian  geometry of the solution \eqref{eq:dilsol}.    In  global coordinates, the  Lorenzian dilaton profile is given by 
 \be
 \Phi=\left(\f{\nu \bar{\phi}}{\sin \pi \alpha \nu} \right)\f{\cos \tau}{\cos \mu}+\left(\f{2\nu \bar{\phi}}{\tan \pi \alpha \nu} +2\pi^{2}  G \gamma^{2} \right)\tan \mu -4\pi G \gamma^{2} \left(\mu \tan \mu +1 \right).
 \ee

  It represents  a two sided black hole, with long interior region.  The locations of these  horizons  are  the critical points of the Lorenzian dilaton profile $\nabla_{a}\Phi=0$. The region between two horizons corresponds to the interior of the black hole. As we increase $\gamma$, the interior region gets larger and larger.   In general, the temperatures of the  two horizons are  different.     However,  as we saw in \eqref{eq:tempcoincide}, in the large $\gamma$ limit, two temperatures eventually agree.  Note again this solution  is slightly  different from the one in \cite{Bak:2007jm} by the chioce of the  sourceless part of dilaton.  If we choose the sourceless part as in \cite{Bak:2007jm}, then the temperature of the two black holes always agree. However this does not yield the reparametrization mode which satisfies the boundary conditions \eqref{eq;bdydil}. 

\subsection*{Acknowledgments}

The author thanks Vijay Balasubramanian, G\'abor S\'arosi  for useful discussions. TU was supported by JSPS Grant-in-Aid for Young Scientists  19K14716.  

\newpage 

\bibliographystyle{JHEP}
\bibliography{RRD}

\providecommand{\href}[2]{#2}\begingroup\raggedright\begin{thebibliography}{10}

\bibitem{umegaki1962conditional}
H.~Umegaki, \emph{Conditional expectation in an operator algebra, iv (entropy
  and information)},  in \emph{Kodai Mathematical Seminar Reports}, vol.~14,
  pp.~59--85, Department of Mathematics, Tokyo Institute of Technology, 1962.

\bibitem{Wall:2011hj}
A.~C. Wall, \emph{{A proof of the generalized second law for rapidly changing
  fields and arbitrary horizon slices}},
  \href{https://doi.org/10.1103/PhysRevD.85.104049}{\emph{Phys. Rev. D}
  {\bfseries 85} (2012) 104049}
  [\href{https://arxiv.org/abs/1105.3445}{{\ttfamily 1105.3445}}].

\bibitem{Bousso:2014sda}
R.~Bousso, H.~Casini, Z.~Fisher and J.~Maldacena, \emph{{Proof of a Quantum
  Bousso Bound}}, \href{https://doi.org/10.1103/PhysRevD.90.044002}{\emph{Phys.
  Rev. D} {\bfseries 90} (2014) 044002}
  [\href{https://arxiv.org/abs/1404.5635}{{\ttfamily 1404.5635}}].

\bibitem{Faulkner:2016mzt}
T.~Faulkner, R.~G. Leigh, O.~Parrikar and H.~Wang, \emph{{Modular Hamiltonians
  for Deformed Half-Spaces and the Averaged Null Energy Condition}},
  \href{https://doi.org/10.1007/JHEP09(2016)038}{\emph{JHEP} {\bfseries 09}
  (2016) 038} [\href{https://arxiv.org/abs/1605.08072}{{\ttfamily
  1605.08072}}].

\bibitem{Witten:2018zxz}
E.~Witten, \emph{{APS Medal for Exceptional Achievement in Research: Invited
  article on entanglement properties of quantum field theory}},
  \href{https://doi.org/10.1103/RevModPhys.90.045003}{\emph{Rev. Mod. Phys.}
  {\bfseries 90} (2018) 045003}
  [\href{https://arxiv.org/abs/1803.04993}{{\ttfamily 1803.04993}}].

\bibitem{Faulkner:2014jva}
T.~Faulkner, \emph{{Bulk Emergence and the RG Flow of Entanglement Entropy}},
  \href{https://doi.org/10.1007/JHEP05(2015)033}{\emph{JHEP} {\bfseries 05}
  (2015) 033} [\href{https://arxiv.org/abs/1412.5648}{{\ttfamily 1412.5648}}].

\bibitem{Hollands:2012sf}
S.~Hollands and R.~M. Wald, \emph{{Stability of Black Holes and Black Branes}},
  \href{https://doi.org/10.1007/s00220-012-1638-1}{\emph{Commun. Math. Phys.}
  {\bfseries 321} (2013) 629}
  [\href{https://arxiv.org/abs/1201.0463}{{\ttfamily 1201.0463}}].

\bibitem{Lashkari:2015hha}
N.~Lashkari and M.~Van~Raamsdonk, \emph{{Canonical Energy is Quantum Fisher
  Information}}, \href{https://doi.org/10.1007/JHEP04(2016)153}{\emph{JHEP}
  {\bfseries 04} (2016) 153}
  [\href{https://arxiv.org/abs/1508.00897}{{\ttfamily 1508.00897}}].

\bibitem{Ryu:2006ef}
S.~Ryu and T.~Takayanagi, \emph{{Aspects of Holographic Entanglement Entropy}},
  \href{https://doi.org/10.1088/1126-6708/2006/08/045}{\emph{JHEP} {\bfseries
  08} (2006) 045} [\href{https://arxiv.org/abs/hep-th/0605073}{{\ttfamily
  hep-th/0605073}}].

\bibitem{Ryu:2006bv}
S.~Ryu and T.~Takayanagi, \emph{{Holographic derivation of entanglement entropy
  from AdS/CFT}},
  \href{https://doi.org/10.1103/PhysRevLett.96.181602}{\emph{Phys. Rev. Lett.}
  {\bfseries 96} (2006) 181602}
  [\href{https://arxiv.org/abs/hep-th/0603001}{{\ttfamily hep-th/0603001}}].

\bibitem{Hubeny:2007xt}
V.~E. Hubeny, M.~Rangamani and T.~Takayanagi, \emph{{A Covariant holographic
  entanglement entropy proposal}},
  \href{https://doi.org/10.1088/1126-6708/2007/07/062}{\emph{JHEP} {\bfseries
  07} (2007) 062} [\href{https://arxiv.org/abs/0705.0016}{{\ttfamily
  0705.0016}}].

\bibitem{Sarosi:2016atx}
G.~S\'arosi and T.~Ugajin, \emph{{Relative entropy of excited states in
  conformal field theories of arbitrary dimensions}},
  \href{https://doi.org/10.1007/JHEP02(2017)060}{\emph{JHEP} {\bfseries 02}
  (2017) 060} [\href{https://arxiv.org/abs/1611.02959}{{\ttfamily
  1611.02959}}].

\bibitem{Sarosi:2016oks}
G.~S\'arosi and T.~Ugajin, \emph{{Relative entropy of excited states in two
  dimensional conformal field theories}},
  \href{https://doi.org/10.1007/JHEP07(2016)114}{\emph{JHEP} {\bfseries 07}
  (2016) 114} [\href{https://arxiv.org/abs/1603.03057}{{\ttfamily
  1603.03057}}].

\bibitem{Sarosi:2017rsq}
G.~S\'arosi and T.~Ugajin, \emph{{Modular Hamiltonians of excited states, OPE
  blocks and emergent bulk fields}},
  \href{https://doi.org/10.1007/JHEP01(2018)012}{\emph{JHEP} {\bfseries 01}
  (2018) 012} [\href{https://arxiv.org/abs/1705.01486}{{\ttfamily
  1705.01486}}].

\bibitem{Faulkner:2017tkh}
T.~Faulkner, F.~M. Haehl, E.~Hijano, O.~Parrikar, C.~Rabideau and
  M.~Van~Raamsdonk, \emph{{Nonlinear Gravity from Entanglement in Conformal
  Field Theories}}, \href{https://doi.org/10.1007/JHEP08(2017)057}{\emph{JHEP}
  {\bfseries 08} (2017) 057}
  [\href{https://arxiv.org/abs/1705.03026}{{\ttfamily 1705.03026}}].

\bibitem{Lashkari:2018oke}
N.~Lashkari, H.~Liu and S.~Rajagopal, \emph{{Modular Flow of Excited States}},
  \href{https://arxiv.org/abs/1811.05052}{{\ttfamily 1811.05052}}.

\bibitem{Bhattacharya:2012mi}
J.~Bhattacharya, M.~Nozaki, T.~Takayanagi and T.~Ugajin, \emph{{Thermodynamical
  Property of Entanglement Entropy for Excited States}},
  \href{https://doi.org/10.1103/PhysRevLett.110.091602}{\emph{Phys. Rev. Lett.}
  {\bfseries 110} (2013) 091602}
  [\href{https://arxiv.org/abs/1212.1164}{{\ttfamily 1212.1164}}].

\bibitem{Blanco:2013joa}
D.~D. Blanco, H.~Casini, L.-Y. Hung and R.~C. Myers, \emph{{Relative Entropy
  and Holography}}, \href{https://doi.org/10.1007/JHEP08(2013)060}{\emph{JHEP}
  {\bfseries 08} (2013) 060} [\href{https://arxiv.org/abs/1305.3182}{{\ttfamily
  1305.3182}}].

\bibitem{Wong:2013gua}
G.~Wong, I.~Klich, L.~A. Pando~Zayas and D.~Vaman, \emph{{Entanglement
  Temperature and Entanglement Entropy of Excited States}},
  \href{https://doi.org/10.1007/JHEP12(2013)020}{\emph{JHEP} {\bfseries 12}
  (2013) 020} [\href{https://arxiv.org/abs/1305.3291}{{\ttfamily 1305.3291}}].

\bibitem{Haehl:2017sot}
F.~M. Haehl, E.~Hijano, O.~Parrikar and C.~Rabideau, \emph{{Higher Curvature
  Gravity from Entanglement in Conformal Field Theories}},
  \href{https://doi.org/10.1103/PhysRevLett.120.201602}{\emph{Phys. Rev. Lett.}
  {\bfseries 120} (2018) 201602}
  [\href{https://arxiv.org/abs/1712.06620}{{\ttfamily 1712.06620}}].

\bibitem{Faulkner:2013ica}
T.~Faulkner, M.~Guica, T.~Hartman, R.~C. Myers and M.~Van~Raamsdonk,
  \emph{{Gravitation from Entanglement in Holographic CFTs}},
  \href{https://doi.org/10.1007/JHEP03(2014)051}{\emph{JHEP} {\bfseries 03}
  (2014) 051} [\href{https://arxiv.org/abs/1312.7856}{{\ttfamily 1312.7856}}].

\bibitem{PETZ198657}
D.~Petz, \emph{Quasi-entropies for finite quantum systems},
  \href{https://doi.org/https://doi.org/10.1016/0034-4877(86)90067-4}{\emph{Reports
  on Mathematical Physics} {\bfseries 23} (1986) 57 }.

\bibitem{DBLP:journals/corr/abs-1206-2459}
T.~van Erven and P.~Harremo{\"{e}}s, \emph{R{\'{e}}nyi divergence and
  kullback-leibler divergence}, {\emph{CoRR} {\bfseries abs/1206.2459} (2012) }
  [\href{https://arxiv.org/abs/1206.2459}{{\ttfamily 1206.2459}}].

\bibitem{2015PNAS..112.3275B}
F.~{Brand{\~a}o}, M.~{Horodecki}, N.~{Ng}, J.~{Oppenheim} and S.~{Wehner},
  \emph{{The second laws of quantum thermodynamics}},
  \href{https://doi.org/10.1073/pnas.1411728112}{\emph{Proceedings of the
  National Academy of Science} {\bfseries 112} (2015) 3275}
  [\href{https://arxiv.org/abs/1305.5278}{{\ttfamily 1305.5278}}].

\bibitem{Casini:2018cxg}
H.~Casini, R.~Medina, I.~Salazar~Landea and G.~Torroba, \emph{{Renyi relative
  entropies and renormalization group flows}},
  \href{https://doi.org/10.1007/JHEP09(2018)166}{\emph{JHEP} {\bfseries 09}
  (2018) 166} [\href{https://arxiv.org/abs/1807.03305}{{\ttfamily
  1807.03305}}].

\bibitem{Moosa:2020jwt}
M.~Moosa, P.~Rath and V.~P. Su, \emph{{A Renyi Quantum Null Energy Condition:
  Proof for Free Field Theories}},
  \href{https://arxiv.org/abs/2007.15025}{{\ttfamily 2007.15025}}.

\bibitem{deBoer:2020snb}
J.~de~Boer, V.~Godet, J.~Kastikainen and E.~Keski-Vakkuri, \emph{{Quantum
  hypothesis testing in many-body systems}},
  \href{https://arxiv.org/abs/2007.11711}{{\ttfamily 2007.11711}}.

\bibitem{Lashkari:2018nsl}
N.~Lashkari, \emph{{Constraining Quantum Fields using Modular Theory}},
  \href{https://doi.org/10.1007/JHEP01(2019)059}{\emph{JHEP} {\bfseries 01}
  (2019) 059} [\href{https://arxiv.org/abs/1810.09306}{{\ttfamily
  1810.09306}}].

\bibitem{Ugajin:2018rwd}
T.~Ugajin, \emph{{Perturbative expansions of R\'enyi relative divergences and
  holography}}, \href{https://doi.org/10.1007/JHEP06(2020)053}{\emph{JHEP}
  {\bfseries 06} (2020) 053}
  [\href{https://arxiv.org/abs/1812.01135}{{\ttfamily 1812.01135}}].

\bibitem{Bao:2019aol}
N.~Bao, M.~Moosa and I.~Shehzad, \emph{{The holographic dual of R\'enyi
  relative entropy}},
  \href{https://doi.org/10.1007/JHEP08(2019)099}{\emph{JHEP} {\bfseries 08}
  (2019) 099} [\href{https://arxiv.org/abs/1904.08433}{{\ttfamily
  1904.08433}}].

\bibitem{May:2018tir}
A.~May and E.~Hijano, \emph{{The holographic entropy zoo}},
  \href{https://doi.org/10.1007/JHEP10(2018)036}{\emph{JHEP} {\bfseries 10}
  (2018) 036} [\href{https://arxiv.org/abs/1806.06077}{{\ttfamily
  1806.06077}}].

\bibitem{Bernamonti:2018vmw}
A.~Bernamonti, F.~Galli, R.~C. Myers and J.~Oppenheim, \emph{{Holographic
  second laws of black hole thermodynamics}},
  \href{https://doi.org/10.1007/JHEP07(2018)111}{\emph{JHEP} {\bfseries 07}
  (2018) 111} [\href{https://arxiv.org/abs/1803.03633}{{\ttfamily
  1803.03633}}].

\bibitem{Balakrishnan:2020lbp}
S.~Balakrishnan and O.~Parrikar, \emph{{Modular Hamiltonians for Euclidean Path
  Integral States}},  \href{https://arxiv.org/abs/2002.00018}{{\ttfamily
  2002.00018}}.

\bibitem{Jackiw:1984je}
R.~Jackiw, \emph{{Lower Dimensional Gravity}},
  \href{https://doi.org/10.1016/0550-3213(85)90448-1}{\emph{Nucl. Phys. B}
  {\bfseries 252} (1985) 343}.

\bibitem{TEITELBOIM198341}
C.~Teitelboim, \emph{Gravitation and hamiltonian structure in two spacetime
  dimensions}, {\emph{Physics Letters B} {\bfseries 126} (1983) 41 }.

\bibitem{Almheiri:2014cka}
A.~Almheiri and J.~Polchinski, \emph{{Models of AdS$_{2}$ backreaction and
  holography}}, \href{https://doi.org/10.1007/JHEP11(2015)014}{\emph{JHEP}
  {\bfseries 11} (2015) 014} [\href{https://arxiv.org/abs/1402.6334}{{\ttfamily
  1402.6334}}].

\bibitem{Maldacena:2016upp}
J.~Maldacena, D.~Stanford and Z.~Yang, \emph{{Conformal symmetry and its
  breaking in two dimensional Nearly Anti-de-Sitter space}},
  \href{https://doi.org/10.1093/ptep/ptw124}{\emph{PTEP} {\bfseries 2016}
  (2016) 12C104} [\href{https://arxiv.org/abs/1606.01857}{{\ttfamily
  1606.01857}}].

\bibitem{Sarosi:2017ykf}
G.~S\'arosi, \emph{{AdS$_{2}$ holography and the SYK model}},
  \href{https://doi.org/10.22323/1.323.0001}{\emph{PoS} {\bfseries Modave2017}
  (2018) 001} [\href{https://arxiv.org/abs/1711.08482}{{\ttfamily
  1711.08482}}].

\bibitem{Bak:2018txn}
D.~Bak, C.~Kim and S.-H. Yi, \emph{{Bulk view of teleportation and traversable
  wormholes}}, \href{https://doi.org/10.1007/JHEP08(2018)140}{\emph{JHEP}
  {\bfseries 08} (2018) 140}
  [\href{https://arxiv.org/abs/1805.12349}{{\ttfamily 1805.12349}}].

\bibitem{Bak:2007jm}
D.~Bak, M.~Gutperle and S.~Hirano, \emph{{Three dimensional Janus and
  time-dependent black holes}},
  \href{https://doi.org/10.1088/1126-6708/2007/02/068}{\emph{JHEP} {\bfseries
  02} (2007) 068} [\href{https://arxiv.org/abs/hep-th/0701108}{{\ttfamily
  hep-th/0701108}}].

\bibitem{Bak:2007qw}
D.~Bak, M.~Gutperle and A.~Karch, \emph{{Time dependent black holes and thermal
  equilibration}},
  \href{https://doi.org/10.1088/1126-6708/2007/12/034}{\emph{JHEP} {\bfseries
  12} (2007) 034} [\href{https://arxiv.org/abs/0708.3691}{{\ttfamily
  0708.3691}}].

\bibitem{Maldacena:2016hyu}
J.~Maldacena and D.~Stanford, \emph{{Remarks on the Sachdev-Ye-Kitaev model}},
  \href{https://doi.org/10.1103/PhysRevD.94.106002}{\emph{Phys. Rev. D}
  {\bfseries 94} (2016) 106002}
  [\href{https://arxiv.org/abs/1604.07818}{{\ttfamily 1604.07818}}].

\bibitem{Balasubramanian:2020coy}
V.~Balasubramanian, A.~Kar and T.~Ugajin, \emph{{Entanglement between two
  disjoint universes}},  \href{https://arxiv.org/abs/2008.05274}{{\ttfamily
  2008.05274}}.

\bibitem{Chen:2020tes}
Y.~Chen, V.~Gorbenko and J.~Maldacena, \emph{{Bra-ket wormholes in
  gravitationally prepared states}},
  \href{https://arxiv.org/abs/2007.16091}{{\ttfamily 2007.16091}}.

\bibitem{Garcia-Garcia:2020ttf}
A.~M. Garc\'\i{}a-Garc\'\i{}a and V.~Godet, \emph{{Euclidean wormhole in the
  SYK model}},  \href{https://arxiv.org/abs/2010.11633}{{\ttfamily
  2010.11633}}.

\end{thebibliography}\endgroup

\end{document}